\documentclass[aps,reprint,showpacs,superscriptaddress]{revtex4-1}
\usepackage{graphicx}
\usepackage{amsmath}
\usepackage{amssymb}
\usepackage{amsfonts}
\usepackage{dcolumn}
\usepackage{xcolor}
\usepackage{subfigure}
\usepackage{float}
\usepackage{bm}
\usepackage{amsthm}

\usepackage{color} 

\usepackage{soul}

\usepackage{hyperref}
\hypersetup{
colorlinks=true,final=true,
        linkcolor=blue,
        citecolor=blue,
        filecolor=blue,
        urlcolor=blue,
}

\usepackage{bigints}   
\usepackage{epstopdf}   

\newcommand{\liznx}{LiZn$X$}

\begin{document}

\title{Correlation between electronic polarization and shift current in cubic and hexagonal semiconductors \liznx ($X$ = P, As, Sb)}

\author{Urmimala Dey}
\email{urmimala.dey@durham.ac.uk}
\affiliation{Centre for Materials Physics, Durham University, South Road, Durham, DH1 3LE, United Kingdom}

\author{Jeroen van den Brink}
\affiliation{Leibniz IFW Dresden, Helmholtzstr. 20, Dresden, 01069, Germany}
\affiliation{Dresden Center for Computational Materials Science (DCMS), TU Dresden, Dresden, 01062, Germany}
\affiliation{Institute of Theoretical Physics and W{\"u}rzburg-Dresden Cluster of Excellence {\it ct.qmat}, Technische Universit{\"a}t Dresden, 01062 Dresden, Germany}

\author{Rajyavardhan Ray}
\email{r.ray@bitmesra.ac.in}
\affiliation{Leibniz IFW Dresden, Helmholtzstr. 20, Dresden, 01069, Germany}
\affiliation{Dresden Center for Computational Materials Science (DCMS), TU Dresden, Dresden, 01062, Germany}
\affiliation{Department of Physics, Birla Institute of Technology Mesra, Ranchi, 835215, Jharkhand, India}

\date{\today}

\begin{abstract}
The rectified bulk photovoltaic effect (BPVE) in noncentrosymmetric semiconductors, also called shift current, 
is considered promising for optoelectronic devices,  terahertz emission and possibly solar energy harvesting. 
A clear understanding of the shift current mechanism and search for materials with large shift current is, 
therefore, of immense interest. $ABC$ semiconductors LiZn$X$ ($X$ = N, P, As, and Sb)
can be stabilized in cubic as well as hexagonal morphologies lacking inversion symmetry$-$an ideal platform to investigate the significant contributing factors to shift current, such as the role of structure and chemical species. Using density-functional calculations properly accounting for the electronic bandgaps, the shift current conductivities in LiZn$X$ ($X$ = P, As, Sb) are found to be approximately an order of magnitude larger than the well-known counterparts and peak close
to the maximum solar radiation intensity. Notably, hexagonal LiZnSb shows a peak shift current conductivity of $\sim -75 ~\rm{\rm{\mu}}$A/V$^2$ 
and Glass coefficient of $ -20$ $\times$ 10$^{-8}$ cm/V, 
comparable to the highest predicted values in literature. Our comparative analysis reveals a quantitative relationship between the shift current response and the electronic polarization. These findings not only posit Li-Zn-based $ABC$ 
semiconductors as viable material candidates for potential applications 
but also elucidates key aspects of the structure-BPVE relationship.
\end{abstract}


\maketitle

\section{Introduction} 
The optical response of a material is given by a series of linear and nonlinear
processes~\cite{Bloembergen1996,Shen2003,Boyd2008}. Second order
optical and transport properties have been studied extensively in the
last few decades, revealing an intimate connection between Berry phases and
nonlinear
processes~\cite{Xiao2010RMP,Sipe2000,Sodemann2015PRL,Parker2019,Matsyshyn2019}. In nonmagnetic semiconductors lacking inversion symmetry, second order optical response gives rise to a 
bulk photovoltaic effect (BPVE)~\cite{Kraut1979,vonBaltz1981}
in the form of a rectified current in response to a linearly polarized
light, known as shift current~\cite{Nastos2006,Nastos2010,Morimoto2016}, 
which has wide applications in optoelectronic devices~\cite{Tan2016,Cook2017,Matsyshyn2021} and terahertz emission \cite{Harrel2010,Somma2014,Ghalgaoui2018}.

Shift current has long been considered as a promising alternative to the conventional $p$-$n$ junction based solar cell devices as this bulk response arises because of the real-space shift of charge centers in noncentrosymmetric materials, thereby producing a less dissipative photocurrent of topological origin~\cite{Morimoto2016}, not confined to the 
Shockley-Queisser (SQ) limit of conventional solar
cells~\cite{Choi2009,Yang2010,Ogawa2017}.
However, the photoconversion efficiency of BPVE materials is found to be much below the SQ limit for intermediate to large gap semiconductors so far~\cite{Pusch_2023}.

For potential applications, a clear understanding of the BPVE response on structural details along with new materials with large BPVE response is of paramount importance. 
In this regard, a quantitative comparison between theory and experiments have been carried out for a variety of materials, such as the well-known multiferroic BiFeO$_3$~\cite{Young2012}, ferroelectric BaTiO$_3$~\cite{Pal2021} and its derivative~\cite{Pal2021} and SbSI~\cite{Sotome2019}.
Recently, materials with large shift current have also been predicted~\cite{Sadhukhan2020,Brehm2014,Zhang2019,Julen2020}. Nevertheless, dependence of shift current magnitude on crystal structure and chemical species is still not fully understood.

Shift current is a bulk phenomenon dependent on the average distance moved by the charge carriers during optical transition, the so called shift vector. The shift vector is defined by the difference of Berry connections of initial and final states participating in the optical transition~\cite{Young2012,Young2012BTO}. Most studies on the structure-BPVE relationship hitherto have focused on materials with nonvanishing spontaneous polarization. Shift current response therein is found to be dependent on the bonding character and charge delocalization of the electronic states and polarization~\cite{Tan2016}. Fregoso {\it et al.}~\cite{Fregoso2017} have explicitly shown that the zone-averaged shift vector is directly proportional to the difference in electronic polarizations between the initial and final states involved in the optical transition lying across the in-gap chemical potential in insulators.

In ferroelectrics, although the magnitude of the shift current conductivity (SCC) is not directly related to the
total ferroelectric polarization~\cite{Tan2016,Brehm2014}, 
theoretical and experimental studies~\cite{Nakamura2017,Kim2020} on the ferroelectric charge transfer complex tetrathiafulvalene-$p$-chloranil (TTF-CA) suggest that SCC may, in fact, be related to the electronic part
of the polarization ($P^{\rm el}$), quantified in terms of the Berry phases of the Bloch
bands~\cite{Vanderbilt1993}. Specifically, large shift current response can be generated in the lower-symmetry ferroelectric structure of TTF-CA which also possesses significant $P^{\rm el}$, approximately 20 times larger than the ionic contribution, $P^{\text{ion}}$.

A systematic and quantitative understanding of the relationship between SCC and $P^{\rm el}$, however, is lacking. Moreover, it remains unclear if this correlation can also be extended to piezoelectric materials where polarization can only be induced by external strain.

Here, we address the dependence of SCC on the structure and composition, revealing subtle aspects of the structure-BPVE relationship, by carrying out a systematic and comparative density functional (DF) investigation of the SCC in the $ABC$ semiconductors LiZn$X$ ($X$ = N, P, As, Sb).
Many
members of the $ABC$ semiconductor family can
be synthesized in cubic as well as hexagonal structures,
both of which are noncentrosymmetric. The polymorphism in Li-Zn-based semiconductors of $A^IB^{II}C^V$-type
provides a rather unique opportunity to explore the effects of the crystal structure and chemical species on the SCC.

Among the considered  LiZn$X$ ($X$ = N, P, As, Sb) semiconductors, while the first three members are known
to crystallize in the cubic half-Heusler
phases~\cite{Kuriyama1988,Kuriyama1987}, 
LiZnSb naturally exists in the hexagonal
phase~\cite{Toberer2009}.
Interestingly, LiZnSb is found to exhibit
polytypism {i.e.,} it is also possible to synthesize the cubic
analog of LiZnSb at ambient pressure condition~\cite{White2016},
while, cubic to hexagonal phase transition can be induced 
in LiZnP and LiZnAs by external pressure~\cite{Chopra2018}. 

Recently, the semiconducting cubic LiZn$X$ half-Heuslers
have been identified as potential piezoelectric materials~\cite{Roy2012}, 
and the hexagonal variants
are shown to exhibit spontaneous and switchable electric polarization~\cite{Bennett2012}, making them suitable candidates to probe for nonlinear shift current response. It is important to note that the considered cubic half-Heuslers may also find applications as high-performance thermoelectric, spintronic and energy materials~\cite{Chopra2018,Vikram2019}. On the other hand, some of the hexagonal polymorphs are identified as {\it hyperferroelectrics} with unique dielectric behaviors, which can retain polarization regardless of screening and are thus potentially useful in ultrathin and ultrafast switching devices~\cite{Garrity2014}. 

Interestingly, the largest components of shift current
conductivities in the LiZn$X$ semiconductors ($X$ = P, As, Sb) range from $\sim -30 ~\rm{\mu}$A/V$^2$ to $\sim -75 ~\rm{\mu}$A/V$^2$, which are two orders of magnitude larger than that in BiFeO$_3$~\cite{Young2012} and comparable to the highest known photoconductivity values reported in literature~\cite{Brehm2014,Zhang2019,Julen2020}. 
Moreover, all these compounds have bandgaps in the visible and near-infrared regions with a sizable photoconductivity in the visible spectrum, which make them promising
candidates for photovoltaic applications. To ascertain the photoresponse, we also compute the Glass coefficients (GCs) for the 
polar hexagonal structures and find values comparable to the largest predicted values so far, rendering them viable for possible solar energy harvesting device applications.

Our DF calculations reveal a quantitative correlation between $P^{\rm el}$ and SCC in ferroelectric hexagonal LiZn$X$ ($X$ = P, As, Sb). Remarkably, this correlation extends to piezoelectric cubic LiZn$X$ ($X$ = P, As, Sb) compounds as well where the induced electronic polarization can, in principle, be used as a figure of merit for prediction of large SCC response.
Our comparative analysis of the electronic
and optical properties of the considered compounds, therefore, elucidates key factors governing the shift current response in LiZn$X$ semiconductors, in principle extendable to inversion-broken materials in general.

\section{Computational Details}
\label{sec:comp}
We performed DF calculations to study the electronic and
optical properties of LiZn$X$ ($X$ = N, P, As, and Sb) semiconductors
employing the Full-Potential Linearized Augmented Plane Wave (FP-LAPW)
method as implemented in WIEN2k~\cite{WIEN2k2002}. To obtain bandgaps which are comparable 
with their experimental counterparts, Tran-Blaha version of
modified Becke Johnson (TB-mBJ) exchange-correlation
potential~\cite{TBmBJ2009} was considered along with the generalized
gradient approximation (GGA) of Perdew-Burke-Ernzerhof
(PBE)~\cite{PBE1996}.
We included the spin-orbit coupling (SOC) effects, as implemented in WIEN2k, only for LiZnAs and
LiZnSb. Starting with the lattice parameters and atomic
positions from available literature (presumably obtained within GGA using WIEN2k with default force threshold)~\cite{Chopra2018}, we reoptimized the internal parameters, utilizing the $P6_3mc$ space group symmetry such that the force on each atom was less than 1 meV/\AA. For this we used the Full-Potential Local-Orbital (FPLO)
code~\cite{FPLO,fplo_web}. The structures thus obtained were used for further calculations. For the cubic compounds, on the other hand, geometry relaxation was not required as all the atoms are located at the high symmetry positions.

Self-consistent calculations were performed with a $20 \times 20 \times
20$ $k$-mesh grid in the full Brillouin zone (BZ) for the cubic compounds, whereas a $20 \times 20 \times
10$ $k$-mesh was used for the hexagonal analogs. Modified tetrahedron-method of Bl{\"o}chl was employed for
$k$-space integration~\cite{Blochl1994}. The energy and
charge density convergence criteria, respectively, were set to $10^{-5}$ Ry and
$10^{-4}$ e/a.u.$^3$ per unit cell. We chose muffin tin radii ($R_{\rm MT}$) 
of 2.0 a.u. and 2.2 a.u. for Li and Zn atoms, respectively. The $R_{\rm MT}$ values 
for the $X$ atoms were set at 2.0 a.u. (P), 2.3 a.u. (As) and 2.5 a.u. (Sb), 
such that $\text{$R_{\rm MT}$} \times k_{\rm{max}}$ was fixed at 7.0, where,  $k_{\rm{max}}$ is the largest plane wave vector. 

Linear optical response of cubic (hexagonal) LiZn$X$ was calculated on
a $50 \times 50 \times 50$ ($64 \times 64 \times 32$) dense $k$-mesh using
the well-known relations implemented in WIEN2k~\cite{Draxl2006}. The optical 
conductivity $\sigma_{ab} (\omega)$ was obtained from the dielectric tensor  
$\epsilon_{ab} (\omega)$: $\sigma_{ab} = i \omega \epsilon_{ab} (\omega)/4\pi$ $(a, b = x, y, z)$,
where $\hbar \omega$ is the energy of the incident photon. The imaginary part of the dielectric tensor $\epsilon^2_{ab} (\omega)$
can be calculated from the velocity matrix elements as~\cite{Ray2017,Sadhukhan2020} 

\begin{equation}
\begin{split}
\epsilon^2_{ab} (\omega) = &-\frac{4\pi^2e^2}{m^2\omega^2} \bigintssss d\mathbf{k} \sum_{n, l} \big(f [E_{ \mathbf{k}n}] - f [E_{ \mathbf{k}l}] \big) \\
& \times \frac{\langle \mathbf{k}n|\hat{v}_a|\mathbf{k}l\rangle \langle \mathbf{k}l|\hat{v}_b|\mathbf{k}n\rangle}{(E_{ \mathbf{k}n} - E_{ \mathbf{k}l} - \hbar\omega - i\eta)}. 
\end{split}
\label{eqn:dielectric}
\end{equation}
Here, $m$ is the free electron mass, $e$ is the electronic charge, and $\hat{v}_a$ and $\hat{v}_b$ are the velocity operators. $|\mathbf{k}n\rangle$ and $|\mathbf{k}l\rangle$ are the electronic wavefunctions with energy eigenvalues $E_{ \mathbf{k}n}$ and $E_{ \mathbf{k}l}$, respectively, defined at the same crystal momentum $\mathbf{k}$ for direct allowed transitions. $f [E_{ \mathbf{k}n}]$ denotes the Fermi function at energy $E_{ \mathbf{k}n}$. $n$ and $l$ are the band indices and $\eta$ is the broadening parameter. 

The corresponding real part $\epsilon^1_{ab} (\omega)$ can be computed from the Kramer-Kronig relation as~\cite{Draxl2006}
\begin{equation}
\epsilon^1_{ab} (\omega) = \delta_{ab} + \frac{2}{\pi} \mathcal{P} \bigintssss_{0}^{\infty} \omega^\prime\frac{\epsilon^2_{ab} (\omega^\prime)}{{\omega^\prime}^2 - \omega^2}~d\omega^\prime
\end{equation}
where $\delta_{ab}$ is the Kronecker delta function and $\mathcal{P}$ refers to the principal value of the integral. The absorption coefficient was then calculated from the real and imaginary parts of the dielectric tensor~\cite{Draxl2006}
\begin{equation}
 \alpha_{aa} (\omega) = \frac{\sqrt{2}\omega}{c} \bigg(\sqrt{{[\epsilon^1_{aa}(\omega)] }^2 + {[\epsilon^2_{aa}(\omega)] }^2} - \epsilon^1_{aa}(\omega) \bigg)^{1/2}\,.
\end{equation}
Similarly, the optical conductivity can also be obtained from the complex dielectric constant~\cite{Ray2017}.

On the other hand, the second-order rectified current density is given by $j^c (0) = \sigma^c_{ab} (0;
\omega, -\omega) E_a (\omega)  E_b (-\omega)$ $(a, b, c = x, y, z)$, where the output dc current
response is generated along $c$ due to the ac electric fields with frequencies $\omega$ and
$-\omega$ along $a$ and $b$ directions, respectively. In general, the third-rank
conductivity tensor $\sigma^c_{ab} (0; \omega, -\omega)$ is a complex quantity; however, under a
linearly polarized light the response is purely driven by the real part of $\sigma^c_{ab} (0;
\omega, -\omega)$~\cite{Matsyshyn2019}. 

We used the Berry module of the Wannier90 code~\cite{Wannier90,Azpiroz2018} to calculate SCC, which is based on the length gauge formalism introduced by Sipe and Shkrebtii for determining the shift current response for insulators in the independent-particle approximation~\cite{Sipe2000}. In this formalism, zero frequency (dc) SCC response generated in the direction $c$ due to the ac electric fields with frequencies $\omega$ and $-\omega$ along $a$ and $b$ directions (i.e. due to $ab$-polarization of light) is expressed as~\cite{Azpiroz2018}
\begin{equation}
\begin{split}
    \sigma^c_{ab} (0; \omega, -\omega) \\
    = &-\frac{i\pi e^3}{4} \bigintssss d\mathbf{k}  \sum_{n,~l} \big(f [E_{ \mathbf{k}n}] - f [E_{ \mathbf{k}l}] \big) \\
    &\times \Big[r^{a}_{\mathbf{k}ln}r^{b;c}_{\mathbf{k}nl} + r^{b}_{\mathbf{k}ln}r^{a;c}_{\mathbf{k}nl}  \Big]\\
    & \times \Big[\delta(E_{\mathbf{k}l} - E_{ \mathbf{k}n} - \hbar\omega)+\delta(E_{\mathbf{k}n} - E_{ \mathbf{k}l} - \hbar\omega)\Big]
\end{split}
\label{shiftsigma}
\end{equation}
The $k$-space integration for both linear and SCC calculations is performed over the first BZ employing the modified tetrahedron-method of Bl{\"o}chl~\cite{Blochl1994} with the integral measure $d\mathbf{k} = \frac{d^dk}{{(2\pi)}^d}$ in $d$ dimensions.

The dipole matrix elements $r^{a}_{\mathbf{k}nl}$ and the generalized derivative $r^{a;b}_{\mathbf{k}nl}$ are related to the Berry connection of the Bloch bands $A^a_{\mathbf{k}nl}$ as
\begin{eqnarray}
    r^{a}_{\mathbf{k}nl} &=& (1-\delta_{nl})A^a_{\mathbf{k}nl}\\
    r^{a;c}_{\mathbf{k}nl} &=& \frac{\partial r^{a}_{\mathbf{k}nl}}{\partial k_c}  - i(A^c_{\mathbf{k}nn}-A^c_{\mathbf{k}ll})r^{a}_{\mathbf{k}nl}
\end{eqnarray}
where the Berry connection $A^a_{\mathbf{k}nl}$ is expressed in terms of the periodic part of the Bloch states $|u_{\mathbf{k}n}\rangle$:
\begin{equation}
   A^a_{\mathbf{k}nl} = i\bigg\langle u_{\mathbf{k}n}\bigg|\frac{\partial}{\partial k_a} u_{\mathbf{k}l} \bigg\rangle  
\end{equation}

Wannier90 uses the Gaussian approximation (in the limit of vanishingly small width) to the Dirac delta functions with a small broadening factor $\eta$ to avoid numerical divergences due to near degeneracies in the sum over virtual states~\cite{Wannier90,Azpiroz2018}:
\begin{equation}
    \delta(x) = \lim_{\eta\to 0} \frac{1}{\eta\sqrt{2\pi}} e^{-x^2/2\eta^2}
\end{equation}

To evaluate $\sigma_{ab}^{c}$ [Eq. (\ref{shiftsigma})], we constructed a tight-binding Hamiltonian from the maximally localized
Wannier functions (MLWF) of Li-$2s$, Li-$2p$, Zn-$4s$, Zn-$4p$, $X$-$np$ ($X$ = N, P, As, Sb) and 
$X$-$nd$ ($X$ = P, As, Sb) orbitals employing the Wannier90
code~\cite{Wannier90}. Here, $n$ is the principal quantum number of the outermost valence shell.
In the next step, SCC was evaluated
using the post-processing Berry module of Wannier90~\cite{Azpiroz2018} on
a $100 \times 100 \times 100$ $k$-point grid for the cubic half-Heuslers and a $126 \times 126 \times 64$ $k$-mesh for the hexagonal systems to obtain well-converged SCC values. The broadening parameter was chosen to be $\eta = 0.05$ eV, which was the same as for the linear conductivity calculations. Convergence of SCC values were tested for a cubic system on a $50 \times 50 \times 50$ $k$-mesh for different $R_{\rm MT}$ values. Specifically, for cubic LiZnP, the changes in peak positions were visibly negligible for reduction of $R_{\rm MT}$ values by $\sim 5\%$. Further reducing the $R_{\rm MT}$ for Zn atoms by $\sim 10\%$ changed the peak SCC values by $\lesssim4\%$ without any appreciable change in the peak positions.

Piezoelectric coefficients and spontaneous polarizations were calculated with the BerryPI module~\cite{Berrypi} implemented in WIEN2k with TB-mBJ and TB-mBJ+SOC as applicable. Converged results were obtained with a $15 \times 15 \times 15$ $k$-mesh grid for the cubic compounds and a $15 \times 15 \times 8$ $k$-grid for the hexagonal structures. 

\section{Results and Discussions}
Among the LiZn$A$ family ($A$: pnictogens N \ldots\ Bi), the  
semiconductors LiZn$X$ ($X$= N, P, As and Sb)
can be stabilized in both cubic and hexagonal crystal structures. 
While the Sb
compound naturally exists in both cubic and hexagonal
morphologies depending on the synthesis route~\cite{White2016}, a
cubic to hexagonal structural phase transition can be induced in LiZnP and LiZnAs 
by applying an external hydrostatic pressure of $\sim 12.3$ GPa and $\sim 20.3$ GPa,
respectively~\cite{Chopra2018}. In comparison, a cubic to hexagonal phase transition in LiZnN is not likely due to a large energy barrier between the two phases~\cite{Chopra2018}. Nevertheless, LiZnN has 
distinct electronic properties and serves as a contrast to highlight the importance of chemical effects. The electronic 
properties of LiZnN are, therefore, discussed separately. 
On the other hand, LiZnBi, which naturally exists in the hexagonal phase and can also be stabilized in the cubic structure~\cite{Chopra2018}, is a Dirac semimetal~\cite{Cao2017} and
thus, not considered in our interband shift current response study.

\begin{figure}[ht!]
	\centering
\includegraphics[scale=0.90]{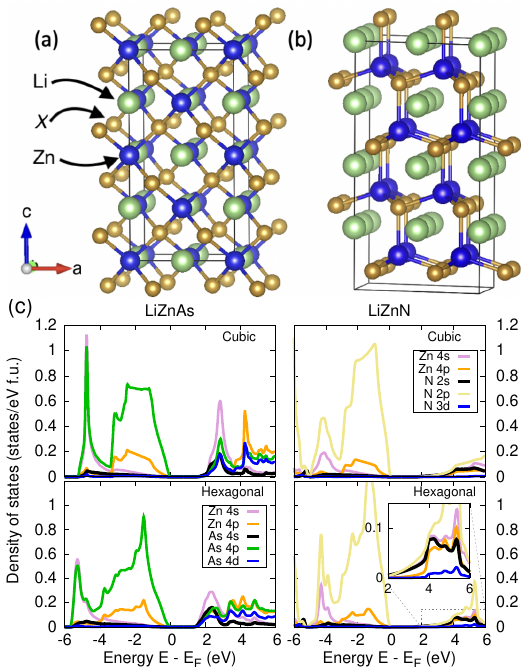}
    \caption{Crystal structures of LiZn$X$ ($X$ = N, P, As and Sb) semiconductors in (a) cubic
   $F\bar{4}3m$ phase and
    (b) hexagonal $P6_3mc$ phase. (c) Elementary electronic properties in terms of selected partial density of states per formula unit for LiZnAs
    (as a representative for $X$ = P, As, Sb) and LiZnN calculated with TB-mBJ potential. The inset shows a zoom-in view of the conduction band edge in hexagonal LiZnN. SOC is included for LiZnAs.}
\label{fig:structures}
\end{figure}
Cubic half-Heusler LiZn$X$ are piezoelectric and crystallize in the $F\bar{4}3m$
(No. 216) MgAgAs-type structure which can be viewed
as a zinc-blende lattice formed by Zn and $X$ atoms with the Li atoms
occupying the tetrahedral interstitial sites~\cite{Wood1985,Kuriyama1991,Kuriyama1996}, as shown
in Fig.~\ref{fig:structures}(a). There are 24 symmetry elements (point group $T_d$) which excludes inversion,
an essential criterion to exhibit second order nonlinear optical effects.

On the other hand, the hexagonal variants are ferroelectrics belonging to the $P6_3mc$ (No. 186)
LiGaGe structure type (point group $C_{6v}$) and possess 12
symmetry operations. Hexagonal LiZn$X$ consists
of a $[$Zn$X]^-$ wurtzite lattice interpenetrated with a Li$^+$
lattice~\cite{Bennett2012,Chopra2018}, as seen from
Fig.~\ref{fig:structures}(b). A polar distortion in the aristotype
$P6_3/mmc$ structure reduces the symmetry to $P6_3mc$ and is associated
with a buckling of the Zn-$X$ planes along $z$ in the wurtzite structure~\cite{Chopra2018,Bennett2012}.

Our starting point is to obtain the optimal crystal
structures for both variants of the considered compounds.
The atomic positions of the resulting structures have
residual forces $\lesssim 1$ meV/{\AA} on each atom, while the lattice parameters
are kept fixed to the values reported in Ref.~\cite{Chopra2018} (see
Sec.~\ref{sec:comp} for details).

In Fig.~\ref{fig:structures}(c), we show the atom- and orbital-resolved density
of states (DOS) per formula unit for LiZnN and LiZnAs as a representative for $X$ = P,
As, and Sb. DOS and electronic band structures of all the LiZn$X$
compounds ($X$ = N, P, As, Sb) are presented in the Supplemental Material (SM)~\cite{supp}.
In both polymorphs of all LiZn$X$ compounds, the whole upper valence
band is dominated by $X$-$np$ states.
For $X$ = P, As, Sb, the lower part of the conduction
band is composed of significant contributions from
Zn-$4s$, Zn-$4p$, $X$-$ns$, $X$-$np$, and $X$-$nd$ states.
The situation for LiZnN is, however, in sharp contrast as no $d$ states exist for $n=2$. In Fig. \ref{fig:structures}(c), 
we, therefore, show the N-$3d$ DOS.
The very large difference between
the N-$3d$ DOS and the X-$nd$ DOS, see Fig.~\ref{fig:structures}(c) and SM~\cite{supp}, is
significant. This arises due to the much higher energy position of
the unoccupied N-$3d$ levels as compared to the $X$-$nd$
levels ($X$ = P, As, Sb). Additionally, the N-$2s$ DOS in the
conduction band is smaller than the As-$4s$ DOS, Fig.~\ref{fig:structures}(c),
and also smaller than the P-$3s$ and Sb-$5s$ DOS (see SM~\cite{supp}).
Together, these have a detrimental consequence for the optical response of LiZnN, as discussed later.

Within the series $X$ = P, As, and Sb, the electronic bandgap
decreases with increasing size of the $X$
ion, as expected. For the cubic structures, the gaps
within GGA are found to be 1.35 eV, 0.41 eV, and 0.33
eV for $X$ = P, As, and Sb, respectively. In comparison,
for the hexagonal structures, the corresponding values
are 1.15 eV, 0.35 eV, and 0.20 eV, respectively. All the
compounds have direct bandgaps at the $\Gamma$ point of the
BZ, except for cubic LiZnP, which is an
indirect bandgap semiconductor with the valence band
maximum at $\Gamma$ and conduction band minimum at the $X$
point~\cite{Chopra2018,supp}.  
The GGA bandgaps found for LiZnN amount to 0.54 eV and 0.35
eV, respectively, for the cubic and hexagonal structures~\cite{supp}.
Here again, LiZnN is distinct from the other LiZn$X$.

The bandgaps obtained within GGA for
the cubic as well as the hexagonal structures are in good agreement
with the previous reports~\cite{Chopra2018}. However, comparison of the GGA bandgaps with the available 
experimental bandgaps for the cubic compounds, shows severe underestimation $-$ a well-known issue with semilocal functionals like GGA. Therefore,
in order to obtain bandgaps of cubic LiZn$X$ semiconductors which are comparable
with their experimental counterparts, we use the Tran-Blaha modified
Becke-Johnson (TB-mBJ) potential~\cite{TBmBJ2009} which is a computationally efficient way to address this issue. Indeed, the resulting
bandgaps are in excellent agreement with the experimental values,
typically within $\lesssim 2\%$ for $X$ = P, As and $\lesssim 9\%$ for $X$ = N. 
Similarly, TB-mBJ gives rise to significant enhancement of bandgaps for the hexagonal structures, by a factor of about $1.7-4.2$~\cite{supp}. 
Since TB-mBJ leads to excellent agreement between the experimental and DF bandgaps for the cubic compounds, 
the obtained bandgaps for the hexagonal phases are expected to match well with the future experiments. 

\begin{figure*}[ht]
\includegraphics[scale=1.76]{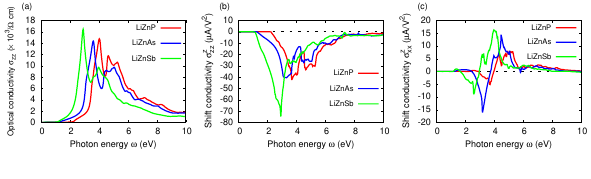}
\caption{(a) $zz$ component of linear optical conductivity; (b), (c) $zzz$ and $zxx$ components of SCC calculated with TB-mBJ potential for hexagonal ferroelectric LiZn$X$ ($X$ = P, As, Sb). SOC is considered for the As and Sb compounds.}
\label{optics_hex}
\twocolumngrid
\end{figure*}

Corrections from the TB-mBJ potential, however, do
not change the nature of the bandgaps, {i.e.,} all the materials remain direct bandgap semiconductors, except for
cubic LiZnP which retains its indirect bandgap.
Moreover, inclusion of TB-mBJ corrections in both phases
produces bandgap values in the visible and near-infrared regions 
of the electromagnetic spectrum, indicating the potential of LiZn$X$ compounds in
photovoltaic applications.
In the following, we will, therefore, evaluate the linear
and nonlinear optical responses of the considered systems
using TB-mBJ.
\begin{figure}[b]
\includegraphics[scale=1.0]{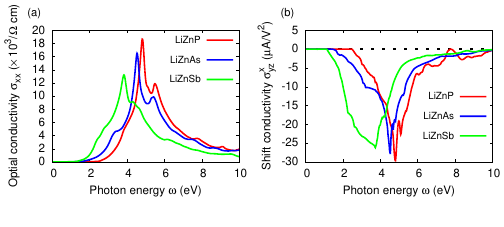}
	\caption{(a) Linear optical conductivity and (b) nonlinear SCC of cubic
    LiZn$X$ ($X$ = P, As, Sb) calculated with TB-mBJ potential. SOC is taken into account for the As and Sb compounds.}
\label{optics_cub}
\end{figure}
Figs.~\ref{optics_cub}(a) and \ref{optics_hex}(a) show the linear
optical conductivity of cubic and hexagonal LiZn$X$ ($X$ = P, As and Sb) semiconductors,
respectively. While only one independent component of $\sigma_{ij}$
appears for the cubic half-Heuslers ($\sigma_{xx} =
\sigma_{yy} = \sigma_{zz}$), the hexagonal symmetry in the ferroelectric
phase allows two independent components ($\sigma_{xx} = \sigma_{yy} \neq
\sigma_{zz}$). For brevity, only the $zz$ component is shown here; the $xx$ component is shown in the SM~\cite{supp}.

The peaks in the linear optical response depend on dipole selection rules via the numerator in Eq. (\ref{eqn:dielectric}) and the joint DOS between the initial and final states. Information about possible bands involved in the (direct) optical transition can be obtained by examining the orbital character of the pair of bands across the Fermi energy satisfying the energy conservation at the high-symmetry points~\cite{Dresselhaus2018}. Qualitative information about the dominant atomic orbital contribution to peaks in optical response can, in principle, be obtained by carefully examining the atom- and orbital-resolved DOS~\cite{Liang1976,Ray2017, Petersen2019}. 
For example, in the cubic compounds, the most prominent peaks at around 4 eV arises from large DOS of As-$4p$ in the valence region and Zn-$4s$ states in the conduction band region (see Fig.~\ref{fig:structures} and SM~\cite{supp}). For the hexagonal compounds as well, the low-energy peaks in the average linear optical conductivity, $\sigma = (2\sigma_{xx} + \sigma_{zz})/3$ (not shown) involve the As-$4p$ states in the valence region, and Zn-$4s$ and As-$4d$ states in the conduction region.

SCC is a third-rank tensor with total 18
components in general. The $T_d$ ($\bar{4}3m$) point group of cubic LiZn$X$ contains three two-fold rotational
symmetries $C_{2x}$, $C_{2y}$ and $C_{2z}$ which invert the sign of all the components of SCC
except when the three indices $x$, $y$ and $z$ appear simultaneously. Thus three non-zero
components survive under the rotational symmetries and the three mirror 
operations $M_{xy}$, $M_{yz}$, $M_{xz}$ further lead to $\sigma^x_{yz} =
\sigma^y_{xz} = \sigma^z_{xy}$. Fig.~\ref{optics_cub}(b) depicts the
variation of $\sigma^x_{yz}$ with the incident photon energy for all the
three cubic LiZn$X$ ($X$ = P, As, Sb) compositions. 
The peak positions of the linear and shift current conductivities agree with each other. This is presumably related
with a peak position of the joint DOS. 

For the hexagonal systems with polar $C_{6v}$ ($6mm$) point group, there is only one two-fold
rotational symmetry $C_{2z}$. Additionally, there are two mirror planes $M_x$ and $M_y$,
perpendicular to the $x$ and $y$ axis, respectively. As a result, the SCC tensor reduces to the
following form:
\begin{equation*}
\big[\sigma^c_{ab}\big]_{\text{hex}} = \begin{pmatrix}
0 & 0 & 0 & 0 & 0 & \sigma^x_{xz} \\
0 & 0 & 0 & 0 & \sigma^y_{yz} & 0 \\
\sigma^z_{xx} & \sigma^z_{yy} & \sigma^z_{zz} & 0 & 0 & 0 \\
\end{pmatrix}
\end{equation*}
Further, the $M_{xy}$ mirror operation leads to $\sigma^x_{xz} =
\sigma^y_{yz}$ and $\sigma^z_{xx} =
\sigma^z_{yy}$, and we have only three independent nonvanishing components
of SCC, among which only the $zzz$ and $zxx$ components are plotted in Figs.~\ref{optics_hex}(b) and ~\ref{optics_hex}(c) for brevity (see SM~\cite{supp} for further details). 

Remarkably, all the LiZn$X$ ($X$ = P, As and Sb) compounds are found to exhibit strong 
shift current response and the magnitude of the
largest component of SCC are $\sim 60-150$ ($\sim 4-10$) times larger than that of the well-known multiferroic
BiFeO$_3$~\cite{Young2012} (SbSI~\cite{Sotome2019}). In particular, in the hexagonal LiZnSb, $\sigma^z_{zz}$ reaches a peak value of $\sim -75
~\rm{\mu}$A/V$^2$, which is comparable to the highest values of SCC predicted for other
materials recently~\cite{Brehm2014,Zhang2019,Julen2020}. 

Evidently, in both cubic and hexagonal phases, the SCC peaks tend to
shift to lower photon frequencies as we move from P to Sb, with hexagonal LiZnSb having the largest $zzz$ component
of SCC in the visible region of the electromagnetic spectrum at photon energy of $\sim$ 2.8
eV, close to the maximum intensity of solar radiation ($\sim 2.5$ eV). 
We emphasize that among the three independent components of SCC in the
hexagonal variants, the $zzz$ components are the largest in magnitude. This can be understood from the
buckling of the Zn-$X$ wurtzite planes along the $z$ direction which is also responsible for spontaneous polarization along $z$.

Shift current arises in noncentrosymmetric materials because of the real-space shift of charge centers under applied electric fields and is of topological origin~\cite{Morimoto2016}. Therefore, to gain further insights into the origin of large shift current in LiZn$X$, we turn our attention to the polarization of these systems. The hexagonal ferroelectrics possess spontaneous polarization while the cubic analogs are piezoelectric, implying that polarization in these systems can be induced by strain. Total polarization (spontaneous or induced) in a noncentrosymmetric material has two different contributions: ionic polarization, $\mathbf{P}^{\text{ion}}$, arising from the displacements of ions and electronic polarization, and $\mathbf{P}^{\text{el}}$, resulting from the Berry phases of the occupied Bloch bands~\cite{Vanderbilt1993}. 

The relationship between total polarization and SCC is generally quite complex. On one hand, it has been established that the magnitude of SCC is not directly related to the
total spontaneous polarization~\cite{Tan2016,Brehm2014} of ferroelectric materials.
On the other hand, it is known that the shift vector, and hence the shift current, is directly proportional to the difference in the Berry connections between the bands participating in the optical transitions~\cite{Young2012,Young2012BTO}. 
Specifically, Fregoso {\it et al.}~\cite{Fregoso2017} have explicitly shown that the zone-averaged shift vector is directly proportional to the difference in electronic polarizations between the initial and final states lying across the Fermi energy in insulators. Based on these, it is argued in Ref.~\cite{Nakamura2017} that materials with large $P^{\rm el}$ would lead to large shift current response. In fact, in systems (e.g. TTF-CA) where bulk polarization is approximately equal to $P^{\rm el}$ ({i.e.,} $P^{\rm el} >> ~P^{\rm ion}$), a controlled dependence of SCC on $P^{\rm el}$ has been demonstrated which is in agreement with the theoretical calculations~\cite{Nakamura2017,Kim2020}.

\begin{figure}[h]
\includegraphics[scale=1.00]{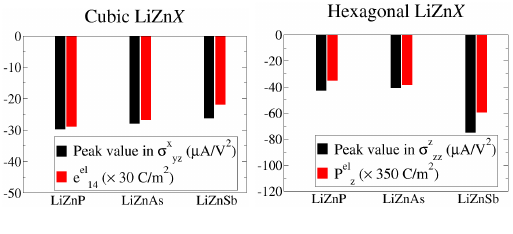}
\caption{Shift current response and degree of inversion symmetry breaking due to optical irradiation, defined in terms of ion-clamped piezoelectric coefficient $e^{\rm el}_{14}$ and electronic polarization $P^{\rm el}_z$ in cubic and hexagonal polymorphs, respectively. All the properties are calculated within TB-mBJ. SOC is included for LiZnAs and LiZnSb.}
\label{prop}
\end{figure}

We calculate both the ionic ($P^{\rm ion}_z$) and electronic ($P^{\rm el}_z$) 
contributions to the spontaneous polarizations for the polar hexagonal polymorphs and find that
while $P^{\rm el}_z$ values are comparable for the P and As compounds, it increases as we move from As
to Sb (Fig.~\ref{prop}). Note that total ferroelectric polarization magnitudes ($P^{\rm tot}_z$) of hexagonal
LiZnAs and LiZnSb~\cite{supp} are in good agreement with previously reported
values~\cite{Bennett2012}, whereas our calculated $P^{\rm tot}_z$ as well as the nature of the bandgap for
hexagonal LiZnP are different from Ref.~\cite{Bennett2012}.  Asymmetry of electronic wavefunctions
forming the covalent bonds give rise to $P^{\rm el}$ which can be considered as a measure of inversion
symmetry breaking due to optical irradiation in polar semiconductors. Therefore, larger $P^{\rm el}_z$ in LiZnSb as compared to LiZnAs
likely originates from relatively larger separation between the positive and negative
charge centers, which in turn leads to larger shift vector and, therefore, larger SCC.

On the other hand, in nonpolar materials, the extent of inversion symmetry breaking due to optical irradiation can be characterized in terms of their piezoelectric response. We compute the ion-clamped piezoelectric coefficient $e^{\rm el}_{14}$ of cubic LiZn$X$ ($X$ = P, As, Sb), which is defined as the induced electronic polarization in piezoelectric materials in response to applied strain. The obtained values of $e^{\rm el}_{14}$, shown in Fig.~\ref{prop}, decrease with the increasing size of the pnictogen atoms. This results in smaller real-space charge separations, giving rise to smaller photoconductivities while moving from cubic LiZnP to cubic LiZnSb. Fig.~\ref{prop} shows a quantitative correlation between $P^{\rm el}$ and peak values of SCC for ferroelectric (hexagonal) LiZn$X$ compounds. Importantly, this correlation generalizes to piezoelectric (cubic) LiZn$X$ systems as well. As a consequence, in the latter, the electronic component(s) of the strain-induced polarization
can act as a figure of merit for large shift current response. 

While the correlation between $P^{\rm el}$ and the resulting shift current is not yet established analytically, we note that both shift current response as well as the electronic polarization depend on the Berry connections of the Bloch bands, indicating a qualitative correlation between the two quantities. Such a notion is further bolstered by the fact that shift current is sensitive to typical electronic structure details such as the nature of bonding and covalency effects~\cite{Tan2016}, and our numerical results reveal that such a correlation exists in the LiZn$X$ ($X$ = P, As, Sb) family of compounds. Additionally, we find the presence of band-resolved Berry curvature hotspots in the BZ leading to large SCC in LiZn$X$ (see SM~\cite{supp} for details).

The interplay between chemical species and the electronic contributions to the piezoelectric
coefficient and polarization is, however, intricate. For example, in hexagonal LiZnN, even if $P^{\rm el}_z$ ($\sim -0.40$ C/m$^2$) is
larger than that of other structural cousins, SCC turns out to be relatively smaller with a peak value of $\sigma^z_{zz} \sim -9~\rm{\mu}$A/V$^2$ 
at photon energy $\sim$ 8.7 eV~\cite{supp}. 
Optical excitations are expected to be
dominated by $X$-$ns$ and $X$-$nd$ states, as well as Zn-$4s$ states in the lower
part of the conduction band. We note that the $X$ atoms in the
hexagonal (cubic) LiZn$X$ compounds have four Zn and three
(four) Li neighbors. Such large coordination numbers lead
to a relatively large $X$-DOS at the conduction band edge,
including contributions from $X$-$ns$, $X$-$np$, and $X$-$nd$ states.
The exception is LiZnN, since $2d$ states do not exist and the
N-$2s$ DOS is also smaller in comparison with the other $X$-$ns$
due to the fact that the N-$2s$ level lies deeper in energy than any other
$X$-$ns$ level. Thus, the number of dipole-allowed optical
transitions from occupied N-$2p$ states is reduced in comparison
to $X$-$np$ by the lack of appropriate empty states.
The relatively low SCC in LiZnN also presumably arises due to the differences in 
the electronic structure of LiZnN compared to other LiZn$X$ compounds ($X$ = P, As, Sb). 

\begin{figure*}[ht]
\includegraphics[scale=1.76]{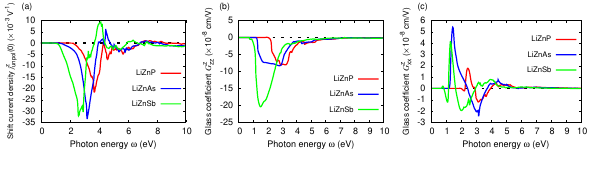}
\caption{(a) Net shift current density per light intensity along $z$ in response to unpolarized light and (b), (c) nonvanishing independent components of Glass coefficient calculated for hexagonal ferroelectric LiZn$X$ ($X$ = P, As, Sb) with TB-mBJ potential. SOC is considered for the As and Sb compounds.}
\label{GC_hex}
\twocolumngrid
\end{figure*}
This suggests that while materials with large electronic part of spontaneous and induced polarizations are ideal candidates 
to probe for the shift current response, one should also look for compositions where dominant contributions to the band edges come from orbitals of different parity.
Further, comparison of the shift current response between cubic and hexagonal polymorphs for a given chemical composition 
(Fig.~\ref{prop}) suggests that the ferroelectric hexagonal variants with spontaneous electronic polarizations are 
better performers compared to their nonpolar cubic analogs.

To ascertain the viability of these materials, specifically hexagonal LiZnSb, for photovoltaic
applications, it is useful to express the shift current $J^c$(0) in terms of the GC 
$G^c_{aa}$: $J^c(0) = G^c_{aa} W I_i$~\cite{ Brehm2014}, where, $a, c = x, y, z$; $W$ is the width of
the exposed sample and $I_i$ is the incident light intensity. This enables evaluation of
attenuation of incident light intensity and nonhomogeneous intensity distribution across the bulk
samples.
GC is related to the linear absorption
coefficient $\alpha_{aa}$ as~\cite{Brehm2014,Tan2016}
\begin{equation}
      G^{c}_{aa} (\omega)  = \frac{2}{c_0\epsilon_0}\frac{\sigma^{c}_{aa}(0; \omega, -\omega)}{\alpha_{aa}(\omega) }
\end{equation}
where, only the components of SCC diagonal in electric fields contribute. Here, $c_0$ is the speed of light in vacuum and $\epsilon_0$ is the vacuum permittivity. It has been shown
that the off-diagonal elements of SCC can not contribute to the total shift current when exposed to
unpolarized light~\cite{Brehm2014,Tan2016}. Therefore, for practical applications in solar cell
devices, one requires a ferroelectric material with nonzero spontaneous polarization, otherwise the total shift
current generated from the off-diagonal field components will sum to zero~\cite{Sturman2021}. Moreover, we can show that the hexagonal LiZn$X$ compounds can only produce a net shift current along the $z$ direction in response to an unpolarized light~\cite{supp}. For an unpolarized light with 45$^{\circ}$ angle of incidence, the net shift current density per light intensity can be expressed as $j^z_{\rm{unpol}}(0) = (3\sigma^z_{xx} + \sigma^z_{zz}) I_0/(c_0\epsilon_0)$, where $I_0$ denotes the intensity of the incident unpolarized light (see SM~\cite{supp} for details). Investigation of Fig.~\ref{GC_hex}(a) reveals that when exposed to an unpolarized light, the net shift current density in hexagonal LiZn$X$ ($X$ = P, As, Sb) can reach a peak value of $ \sim -22 \times 10^{-3}$ V$^{-1}$ to $\sim -35 \times 10^{-3}$ V$^{-1}$ which are almost two orders of magnitude larger than the unpolarized light response in BiFeO$_3$~\cite{Young2012,Yang2010}.
We next compute GC for
the polar hexagonal compounds, as shown in Figs.~\ref{GC_hex}(b) and \ref{GC_hex}(c). From Fig.~\ref{GC_hex}(a), we find that $zzz$ components of GC
for hexagonal LiZn$X$ ($X$ = P, As, Sb) are in the range of $\sim -8 \times 10^{-8}$ cm/V to $\sim -20 \times
10^{-8}$ cm/V, which are $\sim$ 4 $-$ 10 ($\sim$ 2 $-$ 5) times larger than that of BC$_2$N~\cite{Julen2020} (LiAsSe$_2$~\cite{Brehm2014}) and comparable to the largest reported values in literature~\cite{Osterhoudt_2019}. 
Moreover, in all the compounds, the peaks of GC are situated in the visible
range of
the solar spectrum, indicating their use in possible solar energy harvesting devices.

\section{Conclusions and Outlook}
In summary, we have investigated the electronic and optical properties of $ABC$ semiconductors, LiZn$X$ ($X$ = N, P, As and Sb), to elucidate the structure-BPVE
relationship involving the magnitude of SCC, degree of
inversion symmetry breaking and chemical effects. Our comparative study reveals that while
noncentrosymmetric materials with large electronic component of the induced or spontaneous
polarization are suitable candidates for observing large nonlinear optical effects, 
details of the electronic structure also play an important role in determining the magnitude of the response. Specifically, a 
correlation between the shift current conductivity and the electronic part of polarization is found using DF calculations which extends to piezoelectric (cubic) compounds as well.
A more methodical and analytical understanding of the origin of such a correlation is not available at present. However, recent reports, such as in the molecular solid TTF-CA, combined with the above suggest that these quantities are related, albeit in an intricate manner. A quantitative understanding of the influence of crystal structure and composition on this correlation across different materials classes would be an enriching endeavour both theoretically and experimentally.

From a materials perspective, we find that the
LiZn$X$ ($X$ = P, As and Sb) semiconductors exhibit large shift current conductivities, comparable to the highest reported
values in literature and thus may have potential applications in
photovoltaics. Particularly, the polar hexagonal polymorphs with large glass coefficients in the
visible spectrum are promising candidates. 

A significant advantage of the $ABC$ semiconductors considered here is that their ambient structures have
already been synthesized. Especially, 
LiZnSb naturally exists in the polar hexagonal phase. The relative ease of synthesis and stability of cubic and hexagonal
morphologies suggest that these materials are viable from a technological standpoint. Experimental access to these materials under ambient conditions enables direct verification of our predictions.

These findings will likely
fuel further theoretical and experimental studies on these
materials, expedite discovery of unique potential materials and, in turn, development of new-generation devices based on shift current mechanism. Impact of strain,  as a viable means of tuning the ferroelectric polarization~\cite{Kaner_2020,Ebrahimian_2023}, on SCC in the hexagonal LiZn$X$ compounds as well as extending the quantitative correlation between SCC and $P^{\rm el}$ to recently predicted two-dimensional materials with large SCC~\cite{Wang_2019,Mu_2021} should be particularly interesting.

\section*{Acknowledgments}
We thank Dr. Manuel Richter and Prof. D. P. Rai for helpful discussions and Ulrike Nitzsche for technical assistance with the computational resources in IFW Dresden.
We acknowledge financial support from German Forschungsgemeinschaft (DFG, German Research Foundation) via SFB1143 Project No. A05 and under Germany's Excellence Strategy through W{\"u}rzburg-Dresden Cluster of Excellence on Complexity and Topology in Quantum Matter -- {\it ct.qmat} (EXC 2147, Project No. 390858490). U.D. acknowledges financial support from the Leverhulme Trust.


%

\clearpage
\newpage

\onecolumngrid

\section*{SUPPLEMENTAL MATERIAL}
\setcounter{page}{1}
\setcounter{figure}{0}
\setcounter{table}{0}
\setcounter{section}{0}
\renewcommand{\thepage}{S\arabic{page}}
\renewcommand{\thesection}{S\arabic{section}}
\renewcommand{\thetable}{S\arabic{table}}
\renewcommand{\thefigure}{S\arabic{figure}}


Section \ref{sec:str} presents the structural details for the considered systems and tabulated values of the 
bandgaps, while in Section \ref{sec:elprop} we present additional details of the electronic and optical properties for 
$X$=P, As and Sb. In Section \ref{sec:liznn}, we present the detailed electronic and optical properties of LiZnN. Section \ref{sec:unpol} shows the computational details for calculating net shift current density in response to an unpolarized light.


\section{Structural Details}
\label{sec:str}

\begin{table}[h]
  \centering
	\caption{Crystal structures of cubic and hexagonal LiZn$X$ ($X$ = P, As, Sb). Optimized lattice parameters are taken from Ref~\cite{Chopra2018}. In the $F\bar{4}3m$ structure, Zn atoms sit at the origin, Li and $X$ atoms sit at ($\frac{1}{2}$, $\frac{1}{2}$, $\frac{1}{2}$) and ($\frac{1}{4}$, $\frac{1}{4}$, $\frac{1}{4}$), respectively. On the other hand, positions of different types of atoms in the $P6_3mc$ structure are given by Li at ($0$, $0$, $z_{Li}$), Zn at ($\frac{1}{3}$, $\frac{2}{3}$, $z_{Zn}$) and $X$ at ($\frac{1}{3}$, $\frac{2}{3}$, $z_{X}$). $E^{\rm GGA}_{\rm g}$ and $E^{\rm mBJ}_{\rm g}$ are the bandgaps calculated with GGA and TB-mBJ, respectively. $E^{\rm Lit}_{\rm g}$ is the bandgap available in literature. Spin-orbit coupling (SOC) is taken into account for LiZnAs and LiZnSb. Bandgaps calculated with SOC are given in the parenthesis.}
  \label{tab-S1}
	  \begin{tabular*}{0.975\textwidth}{|c| p{0.8cm} p{1.65cm} p{1.65cm} p{1.45cm}|p{0.8cm} p{0.8cm} p{1.65cm}  p{1.65cm} p{1.45cm} p{1.1cm} p{1.1cm} p{1.1cm}|}
    \hline
    $X$ &\multicolumn{4}{c|}{Cubic}&\multicolumn{8}{c|}{Hexagonal} \\
      \hline
		  & $a$ (\AA) & $E^{\rm GGA}_{\rm g}$ (eV)  & $E^{\rm mBJ}_{\rm g}$ (eV) & $E^{\rm Lit}_{\rm g}$ (eV) & $a$ (\AA)  & $c$ (\AA)  & $E^{\rm GGA}_{\rm g}$ (eV) & $E^{\rm mBJ}_{\rm g}$ (eV) & $E^{\rm Lit}_{\rm g}$ (eV) & $z_{Li}$ & $z_{Zn}$ & $z_{X}$\\
    \hline
     P & 5.76 &  1.35 & 1.99& 2.04$^*$~\cite{Kuriyama1988} &4.03& 6.53&1.15& 1.90& 1.19$^\ddagger$~\cite{Chopra2018} &0.9981&0.7783 &0.1615 \\
     As & 5.97&  0.51 (0.41) & 1.58 (1.49)&1.51$^*$~\cite{Kuriyama1994LiZnAs}& 4.18& 6.77&0.40 (0.35)& 1.36 (1.32)& 0.39$^\ddagger$~\cite{Chopra2018}&0.9998&0.7921&0.1752 \\
       Sb& 6.41&  0.54 (0.33) & 1.41 (1.22)& 1.3$^\dagger$~\cite{White2016}&4.46& 7.22&0.38 (0.20)& 0.93 (0.77)& 0.37$^\ddagger$~\cite{Chopra2018} & 0.9996& 0.8303&0.2131 \\
     \hline
	  \multicolumn{13}{l}{ $*$ denotes the experimental bandgaps.} \\
	  \multicolumn{13}{l}{ $\dagger$ and $\ddagger$, respectively, refer to the theoretical values obtained from TB-mBJ and GGA-PBE calculations without SOC.}
\end{tabular*}
\end{table}


\section{Electronic and Optical Properties}
\label{sec:elprop}
\begin{figure}[h]
\centering
\includegraphics[scale=.5]{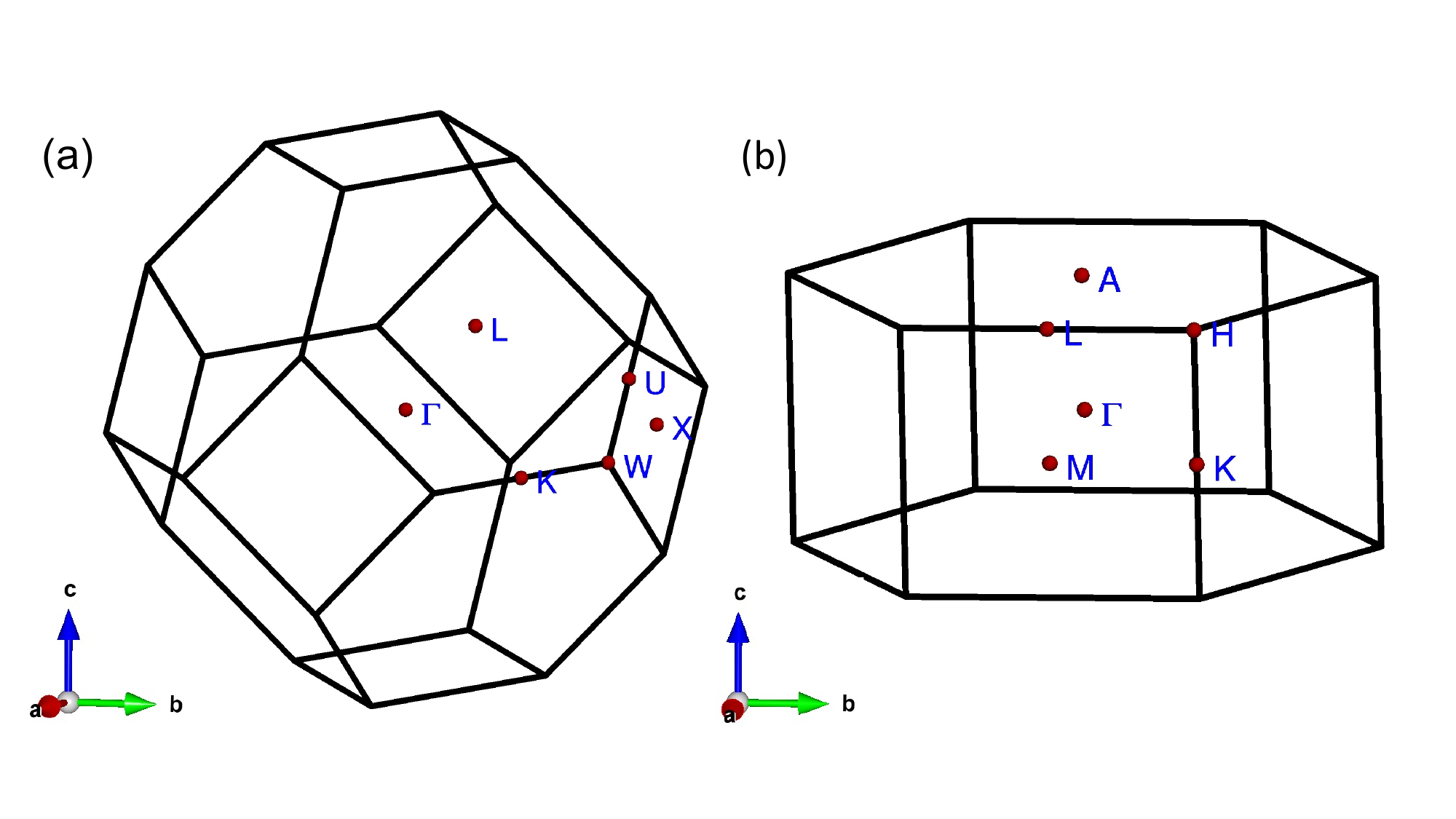}  
\caption{(a) Cubic and (b) hexagonal Brillouin zones with high symmetry points. }
\label{HSP}
\end{figure}
\begin{figure}[h]
\centering
\includegraphics[scale=1.6]{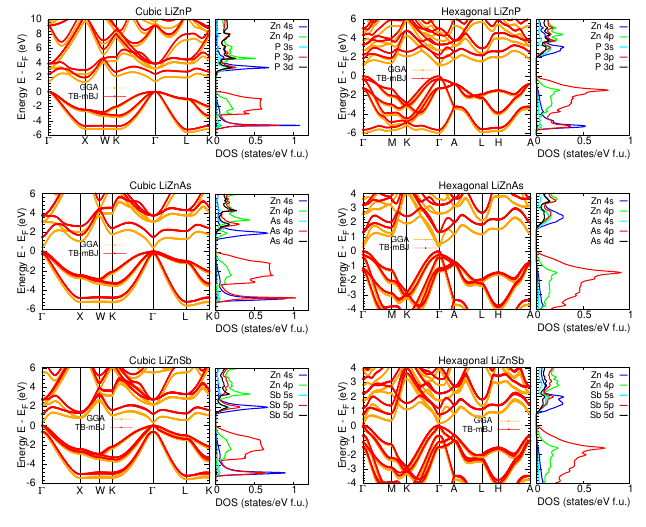}  
\caption{Band structures and density of states (DOS) per formula unit of cubic (left column) and hexagonal (right column) LiZn$X$ ($X$ = P, As, Sb). In each case, band structures calculated with GGA and TB-mBJ are shown in the left and DOS calculated with TB-mBJ are shown in the right. SOC is considered for LiZnAs and LiZnSb.}
\end{figure}

\begin{table}[h]
  \centering
  \caption{Ferroelectric polarization along $z$ calculated for the hexagonal LiZn$X$ ($X$ = P, As, Sb) compounds with TB-mBJ and TB-mBJ+SOC (for LiZnAs and LiZnSb). $P^{\rm ion}_z$ and $P^{\rm el}_z$ denote the ionic and electronic contributions to the spontaneous polarization, respectively. $P^{\rm tot}_z$ is the calculated total contribution. ${[P^{\rm tot}_z]}_{\rm Lit}$ is the literature value of the total contribution calculated within the local density approximation (LDA)~\cite{Bennett2012} and is shown for comparison with our calculated values.}
  \label{tab-S2}
  \begin{tabular}{|c |c| c| c | c |}
    \hline
    Compositions &$P^{\rm ion}_z$ (C/m$^2$)&$P^{\rm el}_z$ (C/m$^2$)& $P^{\rm tot}_z$ (C/m$^2$)&${[P^{\rm tot}_z]}_{\rm Lit}$ (C/m$^2$)  \\
      \hline
   hex-LiZnP&-0.25&-0.10&-0.35&0.84\\
   hex-LiZnAs&0.81&-0.11&0.70&0.75\\
   hex-LiZnSb&0.76&-0.17&0.59&0.56\\
   \hline
\end{tabular}
\end{table}

\begin{table}[h]
  \centering
  \caption{Piezoelectric coefficient $e_{14}$ of cubic LiZn$X$ ($X$ = P, As, Sb) compounds calculated with TB-mBJ and TB-mBJ+SOC (for LiZnAs and LiZnSb). $e^{\rm ion}_{14}$ and $e^{\rm el}_{14}$ denote the ionic and electronic (ion-clamped) contributions to the piezoelectric coefficient, respectively. $e^{\rm tot}_{14}$ is the calculated total piezoelectric coefficient.  ${[e^{\rm tot}_{14}]}_{\rm Lit}$ is the literature value of the total piezoelectric coefficient calculated within LDA~\cite{Roy2012} and is shown for comparison with our calculated values.}
  \label{tab-S3}
  \begin{tabular}{|c |c| c| c |c|c|}
    \hline
    Compositions &$e^{\rm ion}_{14}$ (C/m$^2$)&$e^{\rm el}_{14}$ (C/m$^2$)& $e^{\rm tot}_{14}$ (C/m$^2$) & ${[e^{\rm tot}_{14}]}_{\rm Lit}$ (C/m$^2$) \\
      \hline
   cub-LiZnP&1.45&-0.96&0.49&0.44\\
   cub-LiZnAs&1.35&-0.89&0.46&0.43\\
   cub-LiZnSb&1.17&-0.73&0.44& $-$\\
     \hline
\end{tabular}
\end{table}

\begin{figure}[h]
\centering
\includegraphics[scale=1.55]{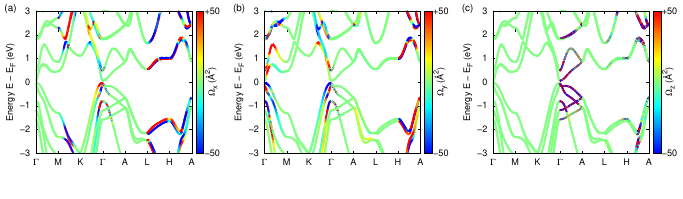}
\caption{Components of Berry curvatures in hexagonal LiZnSb calculated within GGA+SOC framework. The Berry curvature is defined as $\Omega^a_{\mathbf{k}nn} = \epsilon_{abc}\Omega^{bc}_{\mathbf{k}nn} = \frac{\partial}{\partial k_b}A^c_{\mathbf{k}nn} - \frac{\partial }{\partial k_c}A^b_{\mathbf{k}nn}$, where $a, b, c$ are the Cartesian coordinates and $|\mathbf{k}n\rangle$ represents the Bloch state. The high symmetry points correspond to Fig.~\ref{HSP}(b).}
\label{BC}
\end{figure}

\begin{figure}[h]
 \includegraphics[scale=1.5]{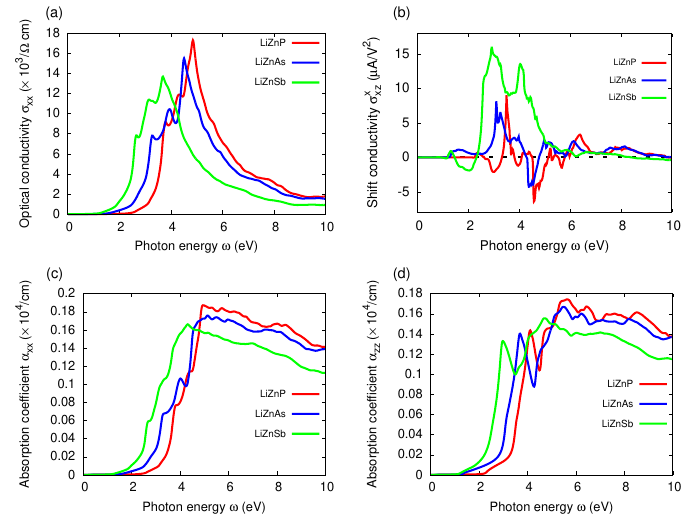}  
\caption{Optical properties of hexagonal LiZn$X$ ($X$ = P, As, Sb): (a) $xx$-component of linear optical conductivity, (b)$xxz$-component of shift current conductivity, (c)$-$(d) linear absorption coefficients.}
\end{figure}
\clearpage
\newpage

\section{ ${\rm{\bf LiZnN}}$ }
\label{sec:liznn}

\begin{table}[h]
  \centering
  \caption{Crystal structures of cubic and hexagonal LiZnN. Optimized lattice parameters are taken from Ref~\cite{Chopra2018}. In the $F\bar{4}3m$ structure, Zn atoms sit at the origin, Li and N atoms sit at ($\frac{1}{2}$, $\frac{1}{2}$, $\frac{1}{2}$) and ($\frac{1}{4}$, $\frac{1}{4}$, $\frac{1}{4}$), respectively. On the other hand, positions of different types of atoms in the $P6_3mc$ structure are given by Li at ($0$, $0$, $z_{Li}$) , Zn at ($\frac{1}{3}$, $\frac{2}{3}$, $z_{Zn}$) and N at ($\frac{1}{3}$, $\frac{2}{3}$, $z_{N}$). $E^{\rm GGA}_{\rm g}$ and $E^{\rm mBJ}_{\rm g}$ are the bandgaps calculated with GGA and TB-mBJ, respectively. $E^{\rm Lit}_{\rm g}$ is the bandgap available in literature.}
  \label{tab-S4}
    \begin{tabular*}{0.9445\textwidth}{|p{0.8cm} p{1.65cm} p{1.65cm} p{1.45cm}|p{0.8cm} p{0.8cm} p{1.65cm}  p{1.65cm} p{1.45cm} p{1.1cm} p{1.1cm} p{1.1cm}|}
    \hline
    \multicolumn{4}{|c|}{Cubic}&\multicolumn{8}{c|}{Hexagonal} \\
      \hline
$a$ (\AA) & $E^{\rm GGA}_{\rm g}$ (eV) & $E^{\rm mBJ}_{\rm g}$ (eV) &$E^{\rm Lit}_{\rm g}$ (eV) & $a$ (\AA)  & $c$ (\AA)  & $E^{\rm GGA}_{\rm g}$ (eV) & $E^{\rm mBJ}_{\rm g}$ (eV) & $E^{\rm Lit}_{\rm g}$ (eV) & $z_{Li}$& $z_{Zn}$ & $z_{N}$\\
    \hline
      4.92&  0.54 & 1.74& 1.91$^*$~\cite{Kuriyama1994}& 3.40& 5.95&0.35&1.46 &0.32$^\ddagger$~\cite{Chopra2018}&-0.0038 &-0.2650&0.09634 \\  
    \hline
    \multicolumn{12}{l}{$*$ denotes the experimental bandgap.} \\
    \multicolumn{12}{l}{$\ddagger$ refers to the theoretical values obtained from GGA-PBE calculations.}
\end{tabular*}
\end{table}

\begin{figure}[h]
\centering
\includegraphics[scale=1.65]{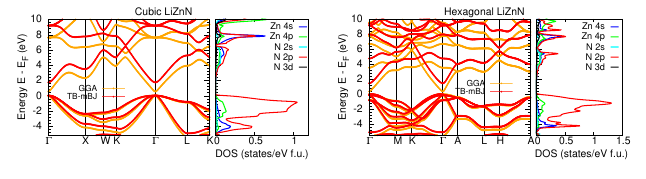}  
\caption{Band structures and density of states (DOS) per formula unit of cubic (left column) and hexagonal (right column) LiZnN. In each case, band structures calculated with GGA and TB-mBJ are shown in the left and DOS calculated with TB-mBJ are shown in the right.}
\end{figure}

\begin{figure}[h]
\centering
\includegraphics[scale=1.54]{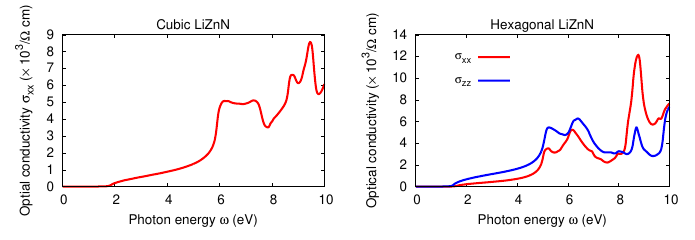}  
\caption{Independent nonvanishing components of linear optical conductivity in cubic (left column) and hexagonal (right column) LiZnN.}
\end{figure}

\begin{figure}[h]
\centering
\includegraphics[scale=1.54]{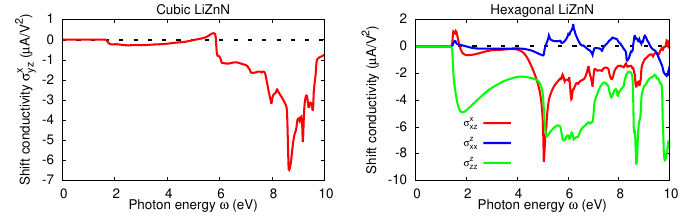}  
\caption{Independent nonvanishing components of shift current conductivity in cubic (left column) and hexagonal (right column) LiZnN.}
\end{figure}
\newpage

\section{Response to unpolarized light}
\label{sec:unpol}

To obtain the shift current response to an unpolarized light, we consider an unpolarized beam incident at an angle to the surface and treat the wavevector components of the electric field propagating in the directions parallel and perpendicular to the interface separately.

Without loss of generality, we consider the wavevector parallel to the interface to be along $y$ axis. We can then resolve the mutually orthogonal polarization components as~\cite{Brehm2014}
\begin{eqnarray*}
    E_1 &=& E_0(\text{cos}\theta ~\hat{i}+ \text{sin}\theta ~\hat{k})\\
    E_2 &=& E_0(\text{-sin}\theta ~\hat{i}+ \text{cos}\theta ~\hat{k})
\end{eqnarray*}
where, $E_0$ is the amplitude of the electric field of the incident unpolarized beam, $\hat{i}$ and $\hat{k}$ are the unit vectors along $x$ and $z$ directions, respectively, and $\theta$ is the angle between the electric field vector $E_1$ and $x$-axis.

Since the off-diagonal elements of SCC do not contribute to the total shift current when exposed to unpolarized light~\cite{Brehm2014}, the cubic non-polar LiZn$X$ compounds will not produce a net current in response to an unpolarized light. On the other hand, in polar hexagonal LiZn$X$, we will have a net shift current density only along the $z$-direction as shown below: 
\begin{eqnarray}
        j^x_{y}(0) &=& 2\sigma^x_{xz}E^x_1E^z_1 + 2\sigma^x_{xz}E^x_2E^z_2 = 0 \nonumber \\
        j^z_{y}(0) &=& \sigma^z_{xx}(E^x_1E^x_1 + E^x_2E^x_2) + \sigma^z_{zz}(E^z_1E^z_1 + E^z_2E^z_2) \nonumber \\
        &=& (\sigma^z_{xx} + \sigma^z_{zz}){(E_0)}^2 \nonumber \\
        &=& \frac{2}{c\epsilon_0}(\sigma^z_{xx} + \sigma^z_{zz}) I_0 \nonumber 
\end{eqnarray}
where $j_{\alpha}^{\beta}(0)$ represents the shift current density along $\beta$ due to incident light with wavevector along $\alpha$, $c$ is the speed of light in vacuum, $\epsilon_0$ is the vacuum permittivity, and $I_0$ is the intensity of the incident unpolarized light. 

Similarly, for the wavevector normal to the interface (propagating along $z$), the net shift current density is given by: 
\begin{eqnarray*}
j^x_{z}(0) &=& 0 \\
j^z_{z}(0) &=& \frac{2}{c\epsilon_0}(\sigma^z_{xx} + \sigma^z_{yy}) I_0 = \frac{4}{c\epsilon_0}\sigma^z_{xx} I_0
\end{eqnarray*}
Therefore, for an unpolarized light with 45$^{\circ}$ angle of incidence, the net current density along $z$ is given by $j^z_{\text{unpol}}(0) = \frac{1}{2}[j^z_{y}(0)+j^z_{z}(0)] = \frac{1}{c\epsilon_0}(3\sigma^z_{xx} + \sigma^z_{zz}) I_0$.

\end{document}